# CONDITIONS OF EXCITATION OF MAGNETOSPHERIC CONVECTION BY THE ELECTRIC CURRENT GENERATED IN THE BOW SHOCK.

E.A.Ponomarev, P.A.Sedykh, O.V.Mager, and V.D.Urbanovich

Institute of Solar-Terrestrial Physics SB RAS, Irkutsk, Russia
pon@iszf.irk.ru , pvlsd@iszf.irk.ru

**Abstract.** This paper analyzes the consequences of electric current generation at the front of the Bow Shock (BS) and the dependence of the direction of this current on the IMF. The conditions of this current closure through the body of the magnetosphere are discussed. It is shown that the process of penetration of the external current into magnetized plasma has a two-stage character. Initially, a change in current on the boundary gives rise to a region of surface charge, the field of which polarizes the near-wall layer with the thickness on the order of one gyroradius of protons. The polarization process involves the formation of the displacement current which produces the Ampere force accelerating the plasma inside the double layer. When the plasma velocity reaches the electric drift velocity (within a time on the order of the inverse gyrofrequency of protons), the electric field in this plasma disappears, whereas in a fixed frame of reference, on the contrary, it reaches equilibrium values. The front of variation of the electric field penetrates the plasma with the velocity of a fast magnetosonic wave. A change in the convection velocity field causes a redistribution of plasma pressure. The appearance of corresponding gradients signifies the penetration of current into plasma. The gradients are changing until a new steady state is reached, to which the new convection velocity field and the new plasma pressure field correspond. This new state is reached in a time $\tau_2$ which is estimated.

## Formulation of the problem

The Bow Shock (BS), separating the Solar Wind (SW) region from the Transition Layer (TL) is, as shown in [Ponomarev et al., 2000], a transformer that converts the solar wind kinetic energy to electric energy. As a result, under the shock wave there arises a current sheet separating the region of the Interplanetary Magnetic Field (IMF) of the solar wind from the magnetic field of the transition layer. If the vertical component of the IMF is in the solar ecliptic coordinate system, $B_z < 0$, then the current under the BS flows in the clockwise direction, and if $B_z > 0$, then it flows in the counterclockwise direction. Fig. 1 shows a portion of the bow shock, the transition layer, and of the fore part of the magnetosphere. The plane of the figure coincides with the XY plane of the magnetospheric coordinate system. For the sake of convenience, we introduced also a local coordinate system **l,n,z**, the axis **l** of which is in opposition to the plasma flow in the TL, and the axis **n** is normal to streamlines, in such a way as to produce a left-handed coordinate system.
All functions or parameters have the index 0 in the SW, 1 in the TL, and 2 in the magnetosphere, so that the mass velocity $v_0$, $v_1$ and $v_2$ implies the plasma velocity in the solar wind, the transition layer, and in the magnetosphere, respectively.

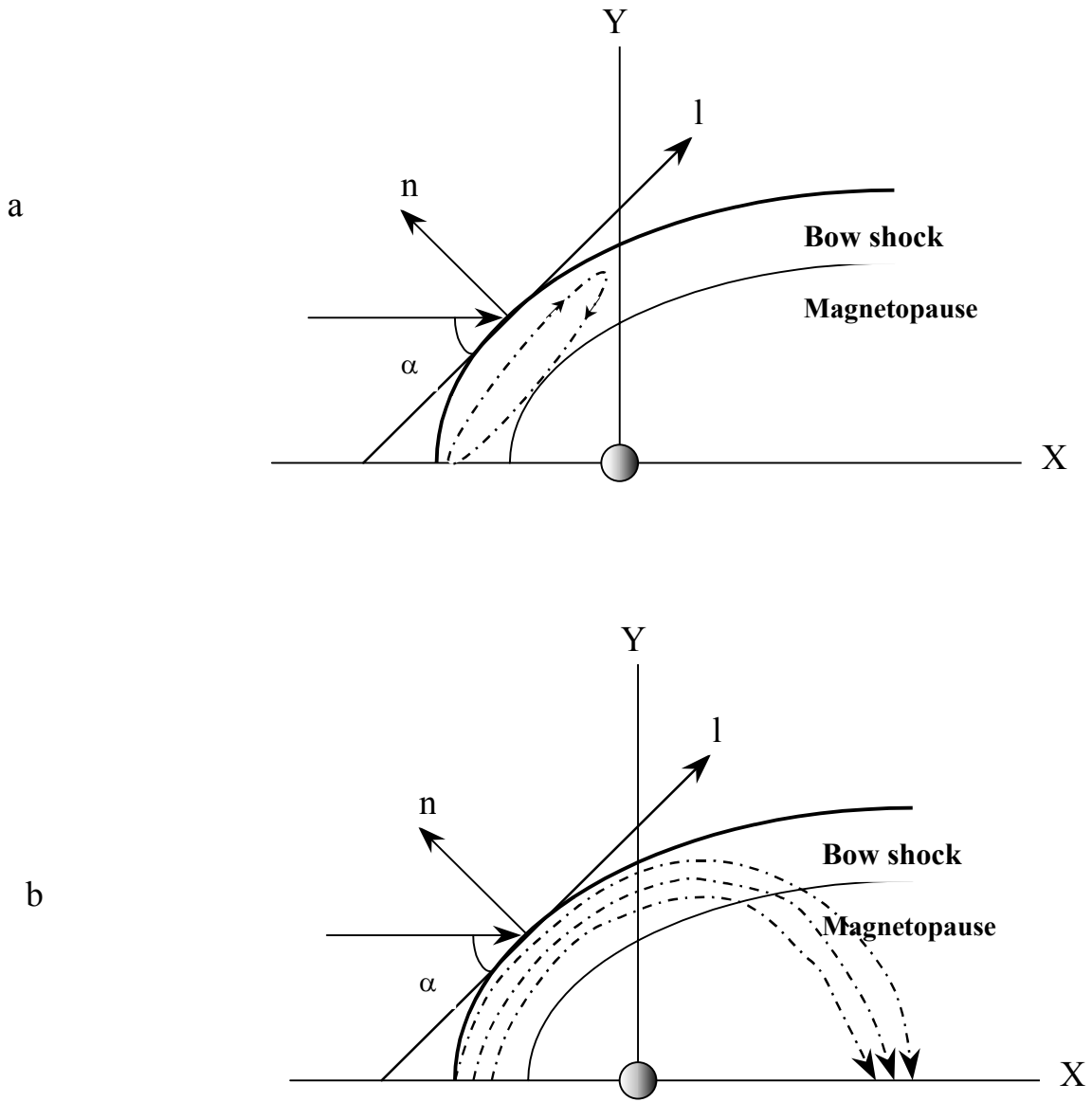

Fig. 1. The position of the bow shock (BS), transition layer and magnetopause (MP). Schematically shown are the density lines of the electric current closing inside the transition layer (a) or through the magnetosphere (b).

We shall approximate the Bow Shock with the hyperboloid of revolution. It is not a very good approximation as regards accuracy, but then it is a very simple one.
Thus:

$$y = [k^2(x-c)^2 - b^2]^{1/2}, \qquad (1)$$

where $r = [x^2 + g^2]^{1/2}$, $g^2 = y^2 + z^2$, $k = b/a$, $c^2 = a^2 + b^2$; a and b being the semiaxes of the hyperboloid with the origin at the focus.

These parameters are conveniently expressed in terms of the distance to the "nose" point, $r_o = x_o$, and $r_c = b^2/a$, the distance from the focus to the surface of the hyperboloid along a normal. In dimensional quantities, which are italicized in the present case, $a = a/x_o$, or example:

$$a = 1/[r_c - 2], \quad b = r_c^{1/2}/[r_c - 2]^{1/2}, \quad c = (r_c-1)/(r_c - 2), \qquad (2)$$

Let $\alpha$ represent the angle between the axis x and a tangent to the surface of the hyperboloid. Then $dr/dx = tg\alpha$. The angle $\alpha$ varies from $\pi/2$ to a limiting value of $\alpha^* = arctg(b/a)$ corresponding to the asymptote. Since $\alpha^*$ is the Mach angle (here we shall use virtually the Alfvén mach):

$$\sin \alpha^* = v_o/V_A = M^{-1}, \qquad (3)$$

the parameters of the hyperboloid can also be expressed in terms of the Mach number.

In some cases it is convenient to use a classical equation for the hyperboloid of revolution:

$$r = y_0/(1 + e \cos\varphi), \qquad (4)$$

where $\varphi$ is the angle between the radius vector directed from the origin to a given point of the surface of the hyperboloid of revolution, $y_0$ is the distance from the origin of coordinates to the surface of the hyperboloid in the plane x=0, $e = (y_0 - x_0)/x_0$ is the eccentricity of the hyperboloid, and $x_0$ is the distance from the origin of coordinates to the nose point of the BS.

In crossing the shock front, the plasma parameters are modified in accordance with the following relations [Korobeinikov et al., 1985]:

$$B_{n0} = B_{n1} \qquad (5)$$
$$\rho_0 v_{n0} = \rho_1 v_{n1} \qquad (6)$$
$$v_{n0} B_{s0} - v_{s0} B_{n0} = v_{n1} B_{s1} - v_{s1} B_{n0} \qquad (7)$$
$$\rho_0 v_{n0}^2 = \rho_0 v_{n0} v_{n1} + p_1 + B_{s1}^2/8\pi \qquad (8)$$
$$\rho_0 v_{n0} v_{s0} - B_{n0} B_{s0}/4\pi = \rho_0 v_{n0} v_{s1} - B_{n0} B_{s1}/4\pi \qquad (9)$$

$$\rho_0 v_{n0} (v_0^2 - v_1^2)/2 = [\gamma p_1/(\gamma-1) + B_{s1}^2/4\pi] v_{n1} \qquad (10)$$

where the indices "n" and "s" designate a normal or tangential field components with respect to the surface of the hyperboloid, the BS front. In (8)-(10) it is taken into consideration that gas and magnetic pressure in the solar wind is much less than dynamic pressure. From (7) and (9) we find:

$$B_{s1} = B_{s0} \sigma [1 - B_{n0}^2/4\pi\rho_0 v_{n0}^2]/[1 - \sigma B_{n0}^2/4\pi\rho_0 v_{n0}^2] \qquad (11)$$

$$v_{s1} = v_{s0} - v_{n0}(\sigma - 1) B_{s0} B_{n0}/4\pi\rho_0 v_{n0}^2 [1 - \sigma B_{n0}^2/4\pi\rho_0 v_{n0}^2] \qquad (12)$$

Here $\sigma$ implies the ratio $(v_{n0}/v_{n1})$. It is evident that with the increasing normal component of the magnetic field, the appearance of a singularity in (11) and (12) may be problem. The infinite increase of the tangential components of the magnetic field and velocity behind the shock front can be avoided if $\sigma$ is made tend to 1. But then, in essence, the shock wave disappears. The transition through the "front" is not accompanied by any jump of velocity, density, etc. Essentially, the problem reduces to seeking the value of the density jump or a normal flow velocity component at the transition through the shock front. Let $\sigma = v_{n0}/v_{n1}$; eliminating $p_1$ from (8) and (10) and using (11) and (12) we then find the equation for $\sigma$:

$$(\sigma-1)(\sigma-\sigma^*)(M_n^2-\sigma)^2 - \sigma^2(\sigma-1)(M_n/M_s)^2 - 2\mathrm{ctg}\alpha \cdot \sigma^2(\sigma-1)(M_n^2-\sigma)(M_n/M_s) +$$
$$+ (2-\gamma)/(\gamma-1) \cdot [\sigma^3 (M_n^2-1)^2]/M_s^2 = 0 \qquad (13)$$

The following designations are adopted here:

$$M_n^2 = (B_0/B_{n0})^2 M^2 \sin^2\alpha, \qquad M_s^2 = (B_0/B_s)^2 M^2 \sin^2\alpha. \qquad (14)$$

$B_0$ is the value of the modulus of the full magnetic field vector of the solar wind; $B_{n0}$ is the value of a normal component of the solar wind magnetic field in a local coordinate system (along the axis **n**); $M_n$, $M_s$ and $M$ are the local mach numbers and the Mach-Alfvén-Mach number, respectively; and $\sigma^* = (\gamma+1)/(\gamma-1)$. The other designations correspond to those adopted previously. It is evident from (13) that at large values of $M_n$ we have $\sigma=\sigma^*$, and when $M_n^2 \Rightarrow \sigma$, the $\sigma$ itself tends to 3! It is remarkable that it is such a value that corresponds to $\sigma^*$ when the adiabatic index $\gamma$ becomes equal to two, as in the case of gas with two degrees of freedom.

**Penetration of the electric current into the magnetosphere and formation of convection**

In a local coordinate system, the magnetic field components are of the form:

$$B_{l1} = \sigma[B_{x0}\cos\alpha + (B_{y0}\sin\upsilon - B_{z0}\cos\upsilon)\sin\alpha] \qquad (15)$$
$$B_{n1} = B_{x0}\sin\alpha - (B_{y0}\sin\upsilon - B_{z0}\cos\upsilon)\cos\alpha \qquad (16)$$
$$B_{\tau 1} = \sigma[B_{y0}\cos\upsilon + B_{z0}\sin\upsilon], \qquad (17)$$

here $\upsilon$ is the angle between the plane $\{x,z\}$ of the magnetospheric coordinate system and the plane $\{l,n\}$ of a local coordinate system. Taking the curl of the magnetic field in a local coordinate system we obtain the value of the current density in this system. In this case our concern is only with the normal component:

$$j_{n1} = c/4\pi [\partial B_{\tau 1}/\partial l - \partial B_{l1}/\partial \tau] \qquad (18)$$

Since $B_{\tau 1}$ is independent of the variables related to the l-coordinate (to $\alpha$, for example), the first term in square brackets in (18) is zero.
Since $d\tau = gd\upsilon$, for the normal current density we have:

$$j_{n1} = -(c\sigma/4\pi g)\cdot[B_{y0}\cos\upsilon + B_{z0}\sin\upsilon]\sin\alpha, \qquad (19)$$

where $g = (y^2 + z^2)^{1/2}$ is defined by equation (1).
Thus we have obtained the expression for the density of the electric field which is produced under the Bow Shock and is directed into the cavity formed by the BS. According to all indications, it is just the current that feeds electromagnetic energy to magnetospheric processes. Firstly, it depends on the $B_y$- and $B_z$-components of the IMF, and, secondly, it has a negative dawn-dusk component that depends only on $B_z$, and the north-southward component that depends only on $B_y$. It is such an association with IMF $B_y$ that is revealed by geomagnetic variations [Uvarov et al., 1989]. Thirdly, a numerical estimation of $j_{n1}$ gives for $\sigma = 4$, $g = 10^{10}$ cm and $B_0 \sim 5$ nT the value of the current density corresponding to the value of the dawn-dusk current density in empirical models [Mead et al., 1975]. At first glance the expression (19) has a nonremovable singularity when $g \to 0$, that is, at the nose point. In fact, this is not the case. Formula (19) describes two streams at a time: one outflowing stream at an angle $\pi/2 - \alpha$ and one inflowing stream at an angle $-(\pi/2 - \alpha)$ to the axis x. At a small g the difference of these angles is also small. In the limiting case these oppositely directed currents coincide, and there is no net current. Clearly, when g are small, currents close in the transition layer and do not penetrate into the magnetosphere (Fig. 1a). One has therefore to introduce a certain parameter, the least value of $g_m$, at which formula (19) is applicable for the magnetosphere. The simplest choice is when $g_m = d$, where d is the TL thickness at the nose point.
Before penetrating the magnetosphere, the current $j_{n1}$ must pass through the transition layer and, hence, we have to discuss the question of the interaction of this current with the TL plasma. Essentially, it has the role of loading, to which some of the electric power is conveyed, yet it can also act as the generator operating in series. All depends on the sign of the scalar product on the right-hand side of the equation:

$$\rho v dv/dt + v\nabla p = \mathbf{jE} \qquad (20)$$

If $\mathbf{jE} > 0$, then electromagnetic energy is consumed in the system; if, however, $\mathbf{jE} < 0$, then, on the contrary, the system generates electromagnetic energy. The second term on the left-hand side of the equality represents unequivocally the generator, since $v\nabla p < 0$. The first term represents the energy consumer, because when the TL plasma travels along the interface between the BS front and the magnetopause, its kinetic energy increases. Thus the question as to whether the TL plasma serves as the generator or loading depends on the relationship of the values of the first and second terms on the left-hand side of (20).

In [Ponomarev, 2000] we showed that with some assumptions about the contribution from "fresh" plasma (i.e. which has just penetrated through the front into the TL), the left-hand side of (20) can operate as a generator. Of course, relying on assumptions is always undesirable. For that reason, in this paper we calculated the $j_{n1}$ directly from the values of the transformed magnetic field. With such a statement of the problem, the role of the inertia term is a merely passive one. In the absence of current, plasma is accelerated only due to a decrease in pressure along the TL, and in the presence of current, it is accelerated additionally. In this case kinetic energy buildup does not depend on the direction of current.

On the other hand, it is clear that the magnetosphere cannot receive more current than produced by the BS.

We now briefly discuss the conditions which must be satisfied for the penetration of current into the magnetosphere. Let at the initial time the electric current be homogeneous and directed along the axis y. If at a certain time the current $j_o$ increases by $\delta j$ outside the volume under consideration, then a charge with surface density $\mu = \int \delta j\, dt$ starts to form on its boundary. The resulting electric field E will give rise to a displacement current $j_c = (\varepsilon/4\pi)\cdot\partial E/\partial t$, that creates an Ampere force which is balanced only by the inertial force, because the corresponding pressure gradient has not yet formed:

$$\rho_o\, \partial v/\partial t = [\mathbf{j_c} \times \mathbf{B}]/c \qquad (21)$$

In magnetospheric conditions dielectric permittivity of plasma $\varepsilon = c^2/V_A^2$, then integrating (21) gives:

$$\mathbf{v} = c[\mathbf{E} \times \mathbf{B}]/B^2, \qquad (22)$$

(in view of the fact that the squares of the perturbed quantities can be neglected as having the second order of smallness).

The physical meaning of what has been said above implies that the polarization electric field is produced inside the double layer with a thickness on the order of $\xi = 2\pi c_s/\omega_B = \pi c/\omega_{pp}$, where $c_s$ is the velocity of the fast magnetosonic wave, and $\omega_{pp}$ is the proton plasma frequency (such a thickness is characteristic for current sheets in collisionless laboratory and space plasmas). In the buildup process this field produces a displacement current that forms an Ampere force which accelerates the plasma inside this layer. Whereas in the buildup process the electric field in the plasma coordinate system drops to zero and in the laboratory coordinate system, on the contrary, it increases to $[\mathbf{V}\times\mathbf{B}]/c$.

Because after the relaxation of the field inside the layer the boundary with surface charge has displaced into the plasma (in a time $\tau_1 \sim 2\pi/\omega_B$) to a distance on the order of $\xi$, all that has been described above is repeated. Obviously, this signifies the penetration of the electric field-associated momentum into the plasma with the velocity of sound. Here is how convection starts to form (see Fig. 2).

If there are no losses in plasma and div$\mathbf{V}$=0, then the story ends up with this. The plasma receives the momentum from the electric field, corresponding to the velocity of steady-state convection.

No stationary electric current is produced in this case. But if the volume is bounded by the walls or if it contains an inhomogeneous magnetic field, then primary convection undergoes restructuring, with the possible formation of a pressure gradient, the existence of which implies that an electric current arises in plasma. This does implies the penetration of the external current into plasma.

Next, we consider the formation of a pressure gradient caused by plasma compressibility. From the continuity equation we have:

$$\rho' = -\rho_o \int \text{div } \mathbf{V} \, dt \qquad (23)$$

The integration goes over the entire time of convection formation $\tau_2$. Therefore (23) may be written as:

$$\rho' = \rho_o (V\nabla p_B/p_B)\tau_2 \sim \rho_o(V/L_B)\tau_2 \qquad (24)$$

The perturbation of pressure $p'$ is determined from the equation of state:

$$p' = c_s^2 \rho', \qquad (25)$$

here $c_s$ is the velocity of the fast magnetosonic wave, since magnetic elasticity is of utmost importance in the magnetosphere.

In view of the fact that the velocity of fast magnetosound in the magnetosphere is virtually equal to the Alfvén velocity, we find

$$\nabla p' \sim -\rho_o c_s^2 V\tau_2/L^2 \sim F, \qquad (26)$$

where F is the Ampere force. Since $V \sim F\tau_1/\rho_o$, from (26) it follows that

$$\tau_1\tau_2 = L^2/c_s^2 \qquad (27)$$

This relation relates the size of the system to the settling time of a stationary distribution of pressure, i.e. to the formation of a gas pressure gradient. This is known to determine the current in plasma.

In other words, $\tau_2$ is the characteristic time of penetration of the electric current into the plasma volume. Assuming for $\tau_1 = \omega_B^{-1} \sim 1$ s, $L \sim 10^{10}$ cm, and for $c_s \sim 3\cdot 10^8$ cm/s, then for $\tau_2$ we find the time of ~1000 s. Such a duration of the transient process for the magnetosphere is unobjectionable. It turns out that our system has three characteristic times: $\tau_1 = 2\pi/\omega$ that characterizes the time of electric field buildup "at a point", $\tau_2$ is the time of penetration of the electric field into the volume, and $\tau_3 = L/c_s$ is the settling time of the electric field in the system of the size L. It should be noted that the gas pressure gradient is also produced when the magnetic field is homogeneous but there is a wall that confines the motion of plasma. Furthermore, the expression (27) retains its form, and the time $\tau_2$ acquires an illustrative character. It is simply the filling time of the volume between the wall and the point that is at a certain distance from the wall, with plasma moving with the velocity V toward the wall.

What has been said above suggests an important physical conclusion. The process of penetration of current into plasma is a two-stage one. Initially, the polarization field is produced, which penetrates into plasma "layer by layer". Or, more exactly, the momentum corresponding to this field penetrates into plasma. Here, if the system is inhomogeneous, the flow can redistribute pressure in such a manner that an electric current arises in plasma because of the appearance of gradients. Energetically, this current is necessary for maintaining convection in the inhomogeneous system. In fact, from the relation:

$$\mathbf{V}\nabla p = \mathbf{E}\,\mathbf{j}, \qquad (28)$$

it follows that current is necessary for maintaining flow in an inhomogeneous medium. If (28) is integrated over the volume of the system, we see that in a stationary case the consumed power:

$$W = \int_U \mathbf{V}\nabla p\, dU = \int_U \text{div}\mathbf{S}\, dU = \int_\Sigma \mathbf{S}\, d\Sigma = \int_\Sigma \psi\mathbf{j}\, d\Sigma, \qquad (29)$$

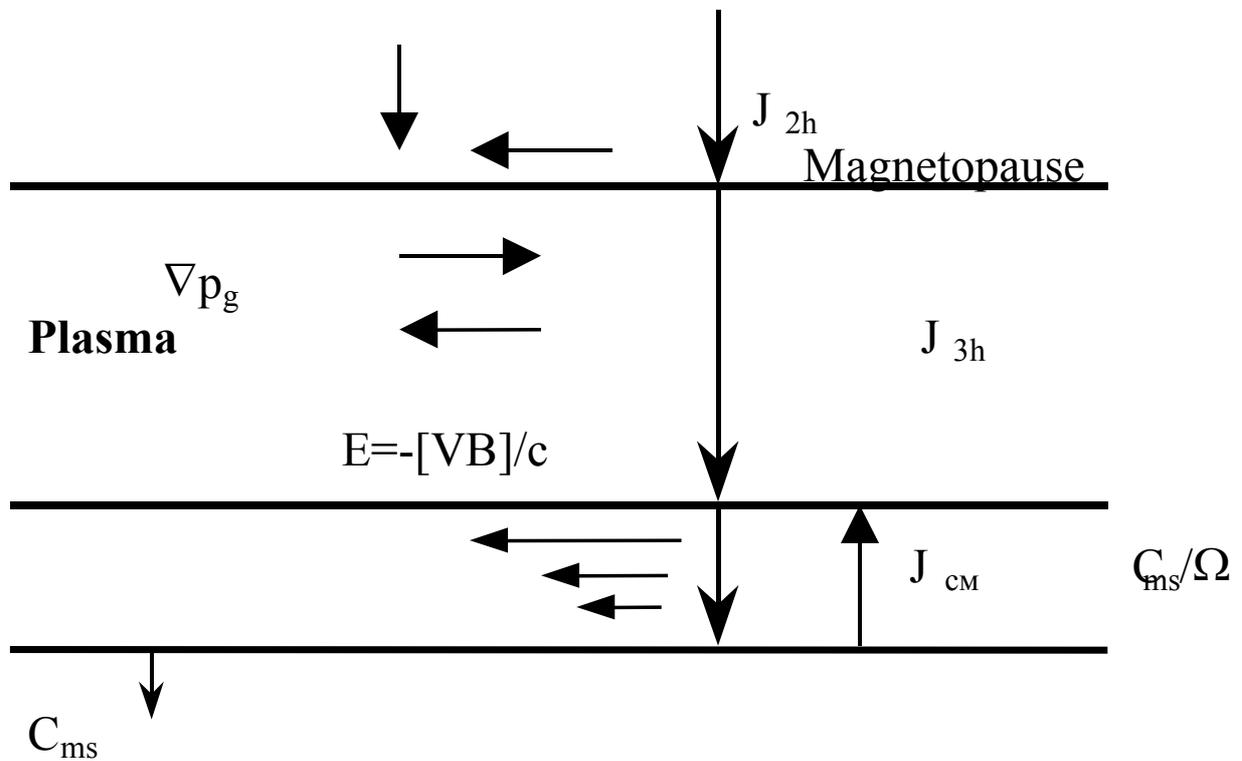

Fig. 2. Schematic illustrating the penetration of the external current into plasma, and excitation of magnetospheric convection.

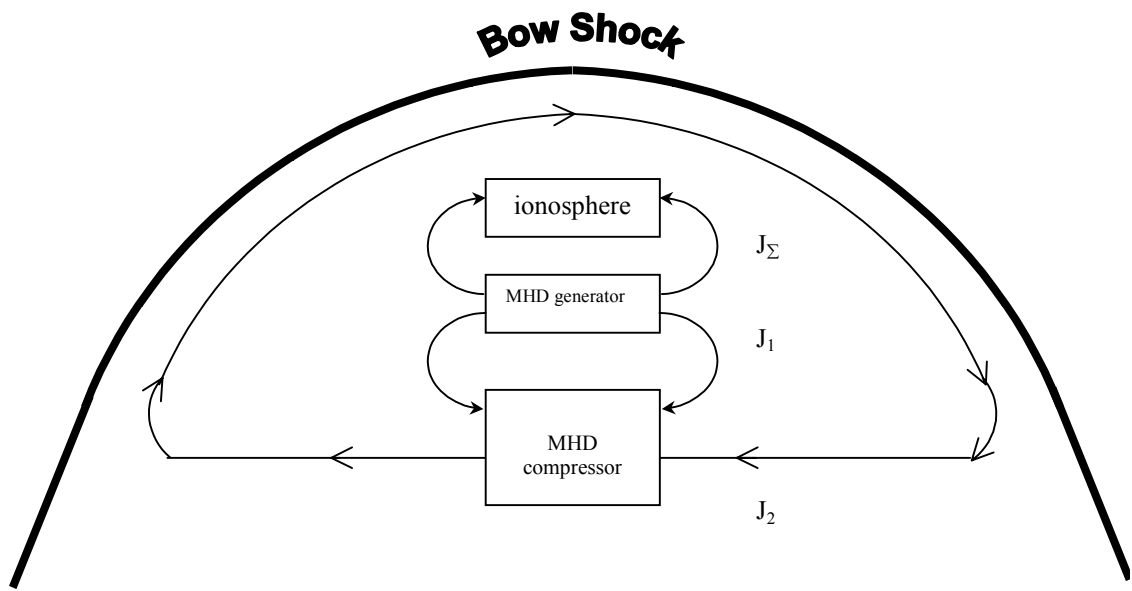

Fig. 3. Layout of the functional blocks in the magnetosphere:
I - MHD generator that converts solar wind kinetic energy to electromagnetic energy;
II - MHD compressor that converts electric energy to gas pressure;
III - secondary MHD generators that convert compressed gas energy to electric current feeding electrojets in the ionosphere.

where U and Σ are the volume and surface area of our system, and Ψ is the surface potential. It is obvious that if the surface is closed and equipotential, i.e. ψ can be taken outside the integral sign, then the system is energetically isolated from its surroundings - no energy can be "pumped in" (or "pumped out"), since $\psi \int_\Sigma \mathbf{j} d\Sigma = 0$. This was demonstrated by Heikkila (1997) . We use (29) to estimate the power brought into the magnetosphere by our current, with an appropriate potential difference. When $\Delta\psi \sim 600$ CGSE ($\sim 120$ keV), $j_{n1} \sim 5 \cdot 10^{-5}$ CGSE and $\Sigma \sim 10^{20}$ cm$^2$, for W we obtain the value on the order of $3 \cdot 10^{18}$ ergs/s, which corresponds to the value adopted for the perturbed state of the magnetosphere.

We have repeatedly stressed that equation (28) has a profound physical meaning. If **Ej**>0, the electric forces do work on plasma. Plasma in this case moves toward increased pressure, that is, it is compressed. Otherwise the expanding gas produces an electric power. In the magnetosphere there are MHD compressors as well as MHD generators (see Fig. 3) [Ponomarev, 2000; Sedykh, Ponomarev, 2002]. If their combined output were identical, then the energy balance would be zero and the magnetosphere would not need any external energy sources. In fact, inside the magnetosphere is a constant energy consumer, the Earth's ionosphere. It is known that the energy flux density through the surface is proportional to the electric field component tangent to this surface.

It is obvious that in this way, by changing the degree of nonequipotentiality of the magnetopause, the magnetosphere can control the energy supplied externally. And the importance of this issue is thus. One can imagine that the intensity of magnetospheric processes is determined by the power alone, which is "offered" by the external source. Yet it may also be assumed that there exists also some regulator that permits the passage only a certain part of the "offered" power. Further, if this regular is linked to the consumer, then we have a very stable natural system.

**Conclusions**

The arguments adduced above suggest the following conclusions.
The bow shock can be a sufficient source of power for supplying energy to substorm processes. The direction of current behind the BS front depends on the sign of the IMF $B_z$-component. When $B_z < 0$ the current in the Bow Shock - magnetosphere circuit flows in the clockwise direction, that is, in the nightside part of the magnetosphere. It is the dawn-dusk current; when $B_z > 0$, it reverses its direction. Any change in external current through the magnetosphere causes a convection restructuring within a time on the order of the travel time of the magnetosonic wave from the magnetopause to the center of the system, because the restructuring wave comes from both flanks. In our model this implies the transition from the two-vortex to four-vortex convection system. Furthermore, the energy possibilities are reduced strongly for the intramagnetospheric processes. Until a new distribution of gas pressure is established, the role of the force counteracting the Ampere force is played by the inertial force. This corresponds to the acceleration of plasma and hence to a change of the electric field. In our opinion, an attempt to observe the restructuring wave in high latitudes with radars of the SuperDARN system or with TV systems with a large angle of view would be worthwhile.

**Acknowledgement.** This work was done under RFBR project №. 02-05-64066, № 03-05-06477.